\documentstyle[12pt]{article}
\newlength{\extraspace}
\newlength{\extraspaces}
\newcommand{\be}{\begin{equation}
\addtolength{\abovedisplayskip}{\extraspaces}
\addtolength{\belowdisplayskip}{\extraspaces}
\addtolength{\abovedisplayshortskip}{\extraspace}
\addtolength{\belowdisplayshortskip}{\extraspace}}
\newcommand{\ee}{\end{equation}}

\newcommand{\ba}{\begin{eqnarray}
\addtolength{\abovedisplayskip}{\extraspaces}
\addtolength{\belowdisplayskip}{\extraspaces}
\addtolength{\abovedisplayshortskip}{\extraspace}
\addtolength{\belowdisplayshortskip}{\extraspace}}
\newcommand{\ea}{\end{eqnarray}}

\newcommand{\nonu}{\nonumber \\[.5mm]}
\newcommand{\A}{&\!\!\!}

\newcommand{\newsection}[1]{
\vspace{7mm} \pagebreak[3] \addtocounter{section}{1}
\setcounter{subsection}{0} \setcounter{footnote}{0}
\begin{center}
{\large {\bf \thesection. #1}}
\end{center}
\nopagebreak
\medskip
\nopagebreak \hspace{3mm}}

\begin{document}

\pagenumbering{arabic}

\begin{center}
{{\bf Regular Charged Solutions in Teleparallel Theory of
Gravity}}
\end{center}
\centerline{ Gamal G.L. Nashed}

\bigskip

\centerline{{\it Mathematics Department, Faculty of Science, Ain
Shams University, Cairo, Egypt }}

\bigskip
 \centerline{ e-mail:nasshed@asunet.shams.eun.eg}

\hspace{2cm}
\\
\\
\\
\\
\\
\\
\\
\\

Using a nonlinear electrodynamics coupled to teleparallel theory
of gravity,  three regular charged spherically symmetric solutions
are obtained. The nonlinear theory reduces to the Maxwell one in
the weak limit and the solutions correspond to  charged
spacetimes. The third solution contains an arbitrary function from
which we can generate the other two solutions.  The metric
associated with these spacetimes is the same, i.e., a regular
charged static spherically symmetric black hole. In calculating
the energy content of the third solution using the gravitational
energy-momentum given by M\o ller, within the framework of the
teleparallel geometry, we find that the resulting form depends on
the arbitrary function. Using the regularized expression of the
gravitational energy-momentum we get the value of energy.

\newpage
\begin{center}
\newsection{\bf Introduction}
\end{center}

Energy-momentum, angular momentum and electric charge play central
roles in modern physics. The conservation of the first two is
related to the homogeneity and isotropy of spacetime respectively
while charge conservation is related to the invariance of the
action integral under internal U(1) transformations. Local
quantities such as energy-momentum, angular momentum and charge
densities are well defined if gravitational fields are not present
in the system. {\it However, in general relativity theory a
well-behaved energy-momentum and angular momentum densities have
not yet been defined, although total energy-momentum and total
angular momentum can be defined for an asymptotically flat
spacetime surrounding an isolated finite system}. The equality of
the gravitational mass and the inertial mass holds within the
framework of general relativity \cite{MTW, SNK,KST}. However, such
equality is not satisfied for the Schwarzschild metric when it is
expressed in a certain coordinate system \cite{BR}.

At present, teleparallel theory seems to be popular again. There
is a trend of analyzing the basic solutions of general relativity
with teleparallel theory and comparing the results.  It is
considered as an essential part of generalized non-Riemannian
theories such as the Poincar$\acute{e}$ gauge theory \cite{Yi1}
$\sim$ \cite{BN} or metric-affine gravity \cite{HMM}. Physics
relevant to geometry may be related to teleparallel description of
gravity \cite{HS1,NH}.  Teleparallel approach is used for
positive-gravitational-energy proof \cite{Me}. A relation between
spinor Lagrangian and teleparallel theory is established
\cite{TN}. Leclerc \cite{Lm5} has shown that the teleparallel
equivalent of general relativity (TEGR) is not consistent in
presence of minimally coupled spinning matter. Mielke \cite{ Me4}
demonstrated the consistency of the coupling of the Dirac fields
to the TEGR. However, Obukhov and Pereira \cite{OP} have shown
that this demonstration is not correct. They also \cite{ OP3} have
studied the general teleparallel gravity model within the
framework of the metric affine gravity theory.

For a satisfactory description of the total energy of an isolated
system it is necessary that the energy-density of the
gravitational field is given in terms of first- and/or
second-order derivatives of the gravitational field variables. It
is well-known that there exists no covariant, nontrivial
expression constructed out of the metric tensor. However,
covariant expressions that contain a quadratic form of first-order
derivatives of the tetrad field are feasible. Thus it is
legitimate to conjecture that the difficulties regarding the
problem of defining the gravitational energy-momentum are related
to the geometrical description of the gravitational field rather
than are an intrinsic drawback of the theory \cite{Mj,MDTC}. M\o
ller has shown that the problem of energy-momentum complex has no
solution in the framework of gravitational field theories based on
Riemannian spacetime \cite{Mo}. In a series of papers,
\cite{Mo}$\sim$\cite{Mo1} he was able to obtain
 a general expression for a satisfactory energy-momentum complex in the teleparallel
  spacetime.

In the last years an increasing revival of nonlinear
electrodynamics  (NLE) theories is observed  \cite{BG}. A
nonlinear electrodynamics was first proposed by Born and Infeld
\cite{BI} at the 30's in order to obtain a finite-energy-electron
model; they succeeded in determining an electron of finite radius.
After these first achievements were carried out, as
Pleb$\acute{a}$nski mentioned in 1970 at the introduction of his
monograph \cite{PP}: {\it If  in recent times the interest in NLE
cannot be said to be very popular, it is not due to the fact that
one could rise some serious objections against this theory. It is
simply rather difficult in its mathematical formulation. What
causes that it is very unlikely to derive some concrete results in
closed form.} It is the aim of the present paper to find
asymptotically flat solutions with spherical symmetry which is
different from Schwarzschild solution in the teleparallel theory
of gravity \cite{Mo1} coupled to nonlinear electrodynamics
\cite{BG}. This can be achieve by inducing the teleparallel
geometry in nonlinear electrodynamics. Applying this philosophy,
we obtain three different exact analytic solutions.  The
singularities of these solutions are studied. We also using the
superpotential of Mikhail et al. \cite{MWHL} to calculate  the
energy.

In \S 2, we briefly review the teleparallel theory of gravitation
coupled  to nonlinear electrodynamics. The tetrad field with three
unknown functions of the radial coordinate in spherical polar
coordinates is applied to the field equations  and three different
exact asymptotically flat solutions are obtained also in \S 3. The
energy of the gravitating source is calculated using the
superpotential method in \S 4. In \S 5 the energy recalculated
using the regularized expression of the gravitational
energy-momentum. Discussion and conclusion of the obtained results
are given in \S 6 \footnote{Computer algebra system Maple 6 is
used in some calculations.}.

\newsection{The tetrad theory of gravitation}

In a spacetime with absolute parallelism the parallel vector
fields ${e_i}^\mu$\footnote{In this paper Latin indices
$(i,j,...)$ represent the vector number, and Greek indices
$(\mu,\nu,...)$ represent the vector components. All indices run
from 0 to 3. The spatial part of Latin indices are denoted by
$(a,b,...)$, while that of Greek indices by $(\alpha,
\beta,...).$} define the nonsymmetric affine connection \be
{\Gamma^\lambda}_{\mu \nu} \stackrel{\rm def.}{=} {e_i}^\lambda
{e^i}_{\ \mu, \ \nu}, \ee where ${e^i}_{\mu, \ \nu}=\partial_\nu
{e^i}_{\mu}$. The curvature tensor defined by
${\Gamma^\lambda}_{\mu \nu}$ is identically vanishing, however.

M\o ller's constructed a gravitational theory based on
 this spacetime. In this
theory the field variables are the 16 tetrad components
${e_i}^\mu$, from which the metric tensor is defined by \be g^{\mu
\nu} \stackrel{\rm def.}{=} \eta^{ij} {e_i}^\mu {e_j}^{\nu}, \ee
where $\eta^{i j}$ is the Minkowski metric $\eta_{i j}=\textrm
{diag}(+1\; ,-1\; ,-1\; ,-1).$

 We note that, associated with any tetrad field ${e_i}^\mu$ there
 is a metric field defined
 uniquely by (2), while a given metric $g^{\mu \nu}$ does not
 determine the tetrad field completely; for any local Lorentz
 transformation of the tetrads ${e_i}^\mu$ leads to a new set of
 tetrads which also satisfy (2).
  The Lagrangian ${\it L}$ is an invariant constructed from
$\gamma_{\mu \nu \rho}$ and $g^{\mu \nu}$, where $\gamma_{\mu \nu
\rho}$ is the contorsion tensor given by \be \gamma_{\mu \nu \rho}
\stackrel{\rm def.}{=} e_{i \ \mu }e_{i \nu; \ \rho}, \ee where
the semicolon denotes covariant differentiation with respect to
Christoffel symbols. The most general Lagrangian density invariant
under the parity operation is given by the form \cite{Mo8} \be
{\cal L} \stackrel{\rm def.}{=} \sqrt{-g} \left( \alpha_1 \Phi^\mu
\Phi_\mu+ \alpha_2 \gamma^{\mu \nu \rho} \gamma_{\mu \nu \rho}+
\alpha_3 \gamma^{\mu \nu \rho} \gamma_{\rho \nu \mu} \right), \ee
where \be g \stackrel{\rm def.}{=} {\rm det}(g_{\mu \nu}),
 \ee
 and
$\Phi_\mu$ is the basic vector field defined by \be \Phi_\mu
\stackrel{\rm def.}{=} {\gamma^\rho}_{\mu \rho}. \ee Here
$\alpha_1, \alpha_2,$ and $\alpha_3$ are constants determined by
M\o ller such that the theory coincides with general relativity in
the weak fields:

\be \alpha_1=-{1 \over \kappa}, \qquad \alpha_2={\lambda \over
\kappa}, \qquad \alpha_3={1 \over \kappa}(1-2\lambda), \ee where
$\kappa$ is the Einstein constant and  $\lambda$ is a free
dimensionless parameter\footnote{Throughout this paper we use the
relativistic units, $c=G=1$ and
 $\kappa=8\pi$.}. The same
choice of the parameters was also obtained by Hayashi and Nakano
\cite{HN}.

The NLE Lagrangian has the form \cite{BG} \be {\cal H}_{NLE}
\stackrel{\rm def.}{=}-\displaystyle{1 \over 4} P_{\mu \nu} P^{\mu
\rho}\; ,
 \ee
 with \ba \A \A P_{\mu \nu}={\cal L(F)}_F F_{\mu \nu}, \quad
{\cal L(F)}_F=\displaystyle {\partial \cal {L(F)} \over \partial
F}, \quad {\cal L(F)}=\displaystyle{1 \over 4} F_{\mu \nu} F^{\mu
\nu} \nonu
\A \A  and \ F_{\mu \nu} \ being \ given \
by\footnote{Heaviside-Lorentz rationalized units will be used.}
 F_{\mu \nu}=
\partial_\mu A_\nu-\partial_\nu A_\mu, \ where \ A_\mu \ is \ the \ vector \
potential.\ea

The gravitational and NLE field equations for the system described
by ${\cal L_G}+{\cal H}_{NLE}$ have the following \cite{BG,Mo8}:
 \ba \A \A G_{\mu \nu} +H_{\mu \nu} =
-{\kappa} T_{\mu \nu}\; , \nonu
 \A \A K_{\mu \nu}=0\; , \nonu
 \A \A \partial_\nu \left( \sqrt{-g} P^{\mu \nu} \right)=0, \ea
where  $G_{\mu \nu}$ is the Einstein tensor, $H_{\mu \nu}$ and
$K_{\mu \nu}$ are defined by \be H_{\mu \nu} \stackrel{\rm
def.}{=} \lambda \left[ \gamma_{\rho \sigma \mu} {\gamma^{\rho
\sigma}}_\nu+\gamma_{\rho \sigma \mu} {\gamma_\nu}^{\rho
\sigma}+\gamma_{\rho \sigma \nu} {\gamma_\mu}^{\rho \sigma}+g_{\mu
\nu} \left( \gamma_{\rho \sigma \tau} \gamma^{\tau \sigma \rho}-{1
\over 2} \gamma_{\rho \sigma \tau} \gamma^{\rho \sigma \tau}
\right) \right],
 \ee
and \be K_{\mu \nu} \stackrel{\rm def.}{=} \lambda \left[
\Phi_{\mu,\nu}-\Phi_{\nu,\mu} -\Phi_\rho \left({\gamma^\rho}_{\mu
\nu}-{\gamma^\rho}_{\nu \mu} \right)+ {{\gamma_{\mu
\nu}}^{\rho}}_{;\rho} \right], \ee and they are symmetric and skew
symmetric tensors, respectively.

M\o ller assumed that the energy-momentum tensor of matter fields
is symmetric. In the Hayashi-Nakano theory, however, the
energy-momentum tensor of spin-$1/2$ fundamental particles has
non-vanishing antisymmetric part arising from the effects due to
intrinsic spin, and the right-hand side of antisymmetric field
equation  (10) does not vanish when we take into account the
possible effects of intrinsic spin.

It can be shown \cite{HS} that the tensors, $H_{\mu \nu}$ and
 $K_{\mu \nu}$, consist of only those terms which are linear or quadratic
in the axial-vector part of the torsion tensor, $a_\mu$, defined
by \be a_\mu \stackrel{\rm def.}{=} {1 \over 3} \epsilon_{\mu \nu
\rho \sigma} \gamma^{\nu \rho \sigma}, \qquad where \qquad
\epsilon_{\mu \nu \rho \sigma} \stackrel{\rm def.}{=} \sqrt{-g}
\delta_{\mu \nu \rho \sigma}, \ee where $\delta_{\mu \nu \rho
\sigma}$ being completely antisymmetric and normalized as
$\delta_{0123}=-1$. Therefore, both $H_{\mu \nu}$ and $F_{\mu
\nu}$ vanish if the $a_\mu$ is vanishing. In other words, when the
$a_\mu$ is found to vanish from the antisymmetric part of the
field equations, (10), the symmetric part of Eq. (10) coincides
with the Einstein field equation in teleparallel equivalent of
general relativity. The energy-momentum tensor $T^{\mu \nu}$ is
defined by \be T^{\mu \nu} \stackrel{\rm def.}{=} 2\left({\cal
H}_P {P^\mu}_{\lambda}P^{\nu \lambda}-\delta^{\mu
\nu}\left[2P{\cal H}_P -{\cal H}\right] \right), \qquad where
\qquad P \stackrel{\rm def.}{=} (1/4)(P_{\mu \nu}P^{\mu \nu}). \ee
\newsection{Spherically Symmetric Solutions}

Let us begin with the tetrad having a spherical symmetry \cite{Ro}

\be \left({e_l}^\mu \right)= \left( \matrix{ A & Dr & 0 & 0
\vspace{3mm} \cr 0 & B \sin\theta \cos\phi & \displaystyle{B \over
r}\cos\theta \cos\phi
 & -\displaystyle{B \sin\phi \over r \sin\theta} \vspace{3mm} \cr
0 & B \sin\theta \sin\phi & \displaystyle{B \over r}\cos\theta
\sin\phi
 & \displaystyle{B \cos\phi \over r \sin\theta} \vspace{3mm} \cr
0 & B \cos\theta & -\displaystyle{B \over r}\sin\theta  & 0 \cr }
\right), \ee where {\it A}, {\it D}, {\it B}, are unknown
functions of $r=(x^\alpha x^\alpha)^{1/2}$. Applying (15) to the
field equations (10) we obtain a system of non linear differential
equations \cite{Ngr}.
 \ba \kappa
T_{0 0} \A= \A {1 \over r A^2 B^4}\Biggl[ \Biggl \{ \Biggl(3
D^2+8B'^2\Biggr) D-2\Biggl(2D B''+B'D'\Biggr)B \Biggr \} r^3 B^2D-
\nonu
\A \A \Biggl \{2\Biggl(D B''+B'D'\Biggr)B-5DB'^2\Biggr \} r^5
D^3-\Biggl(2BB''-3D^2-3B'^2 \Biggr) rB^4+\nonu
\A \A 2\Biggl(BD'-4DB'\Biggr)
r^4BD^3+2\Biggl(BD'-6DB'\Biggr)r^2B^3D-4B^5B' \Biggr], \nonu
\kappa T_{0 1} \A= \A  {D \over  A B^4} \Biggl[ \Biggl \{ 2 \Biggl
(D B''+B'D'\Biggr) B-5DB'^2  \Biggr \}
r^3D+\Biggl(2BB''-3D^2-3B'^2 \Biggr) rB^2-\nonu
\A \A  2\Biggl(B D'-4DB'\Biggr)r^2BD+4B^3B' \Biggr], \nonu
\kappa T_{1 1} \A= \A {1 \over r A B^4}\Biggl[ \Biggl \{ \Biggl(3
D^2+B'^2\Biggr) A+2BA'B' \Biggr \} r B^2- \Biggl \{ 2 \Biggl (D
B''+B'D'\Biggr) B-5DB'^2  \Biggr \} r^3AD + \nonu
\A \A  2\Biggl(B D'-4DB'\Biggr)r^2ABD-2AB^3B'-2B^4A' \Biggr],\nonu
\kappa T_{2 2} \A= \A {r \over  A^2 B^4} \Biggl[ \Biggl ( \Biggl
\{ \Biggl(D A''+3 A'D'\Biggr) B-3DA'B'  \Biggr \} A B D +\Biggl\{
\Biggl (2D B''+5B'D'\Biggr) B D- \nonu
\A \A \Biggl( D D''+D'^2 \Biggr) B^2-5D^2B'^2\Biggr \}
A^2-2B^2D^2A'^2 \Biggr) r^3+\nonu
\A \A \Biggl \{ \Biggl(B'^2 -3D^2\Biggr)
A^2-AB^2A''-B''BA^2+2B^2A'^2 \Biggr\}rB^2-\nonu
\A \A 2\Biggl \{ \Biggl( 3BD'-4DB'\Biggr)A-2BDA' \Biggr \}
r^2ABD+A^2B^3B'+AB^4A' \Biggr ], \nonu
T_{3 3} \A= \A sin \theta^2 T_{2 2}, \ea where
$A'=\displaystyle{dA \over dr}$, $B'=\displaystyle{dB \over dr}$
and  $D'=\displaystyle{dD \over dr}$.

Now we are interested in solving the above differential equations:\\
\underline{Special solutions:}\\
A first non-trivial solution can be obtained by taking $D(r)=0$,
 and solving for $A(r)$ and $B(r)$, then we obtain
 \be A =
\sqrt{1-\displaystyle{2m\left[1-\tanh\left(\displaystyle{q^2 \over
2mR}\right)\right] \over R}}, \qquad \qquad B =
\int{\displaystyle{1 \over R}
 \left(1-\sqrt{1-\displaystyle{2m \over R}\left[1-\tanh\left(\displaystyle{q^2 \over
2mR}\right)\right]}\right)} dR, \ee where $R$ is a new radial
coordinate  defined by $R=r/B$.

The anstaz of the anti-symmetric field $P_{\mu \nu}$, the
nonlinear electrodynamics source used to derive this solution and
the energy-momentum tensor have the form \ba {\bf P} \A =\A
\displaystyle{q \over R^2} dt\wedge dR, \qquad \qquad {\cal
H}=-\displaystyle{q^2 \over 2R^4} sech^2\left(\displaystyle{q^2
\over 2mR}\right), \qquad \qquad {T_0}^0={T_1}^1=\displaystyle{q^2
e^{(q^2/mR)} \over 2\pi R^4\left(1+ e^{(q^2/mR)}\right)^2},\nonu
 {T_2}^2\A =\A{T_3}^3=\displaystyle{q^2e^{(q^2/mR)}\left(q^2(e^{(q^2/mR)}-1)-
 2mR(1+e^{(q^2/mR)}) \right) \over 4 \pi mR^5 (1+e^{(q^2/mR)})^3}.\ea
 Using (17), the tetrad (15) takes the form
\be \left({e_i}^\mu \right)= \left( \matrix{ \displaystyle{1 \over
\sqrt{1- \displaystyle{2m\left[1-\tanh\left({q^2 \over
2mR}\right)\right] \over R}}} &0 & 0 & 0 \vspace{3mm} \cr 0 &
\sin\theta \cos\phi \sqrt{1-
\displaystyle{2m\left[1-\tanh\left({q^2 \over 2mR}\right)\right]
\over R}} & \displaystyle{\cos\theta \cos\phi \over R} &
-\displaystyle{ \sin\phi  \over R \sin\theta} \vspace{3mm} \cr 0 &
\sin\theta \sin\phi \sqrt{1-\displaystyle{2m\left[1-\tanh\left(
 {q^2 \over 2mR}\right)\right] \over R}} &
 \displaystyle{\cos\theta \sin\phi \over R} & \displaystyle{\cos\phi
 \over R \sin\theta} \vspace{3mm}\cr 0 &
\cos\theta\sqrt{1-\displaystyle{2m\left[1-\tanh\left({q^2 \over
2mR}\right)\right] \over R}}& -\displaystyle{\sin\theta \over R} &
0 \cr } \right), \ee with the associated Riemannian metric \be
ds^2=-\eta_1 dt^2+{dR^2 \over \eta_1}+R^2 d\Omega^2, \quad  where
\quad  \eta_1=\left[1-\displaystyle{2m\left[1-\tanh\left({q^2
\over 2mR}\right)\right] \over R} \right] \cong 1-\displaystyle{2m
\over R}+\displaystyle{q^2 \over R^2} , \ee where
${d\Omega^2=d\theta^2+\sin^2\theta d\phi^2}$. Eq. (20) represents
 a static spherically symmetric  regular black hole solution
\cite{BG}

A second non-trivial solution can be obtained by taking $A(r)=1$,
 $B(r)=1$, $D(r)\neq0$ and solving for $D(r)$. In this case the resulting field equations of
 (16) can be integrated directly to give
\be D(r)=\sqrt{2m\left[1-\tanh\left(\displaystyle{q^2 \over
2mr}\right)\right] \over r^3},\ee Substituting for the value of
$D(r)$ as given by (21) into (15), we get
 \be
\left({e_i}^\mu \right)= \left( \matrix{ 1 &
\sqrt{\displaystyle{2m\left[1-\tanh\left({q^2 \over
2mr}\right)\right] \over r}} & 0 & 0 \vspace{3mm} \cr 0 &
\sin\theta \cos\phi & \displaystyle{\cos\theta \cos\phi \over r}
 & -\displaystyle{ \sin\phi \over r \sin\theta} \vspace{3mm} \cr
0 &  \sin\theta \sin\phi & \displaystyle{\cos\theta \sin\phi \over
r} & \displaystyle{\cos\phi \over r \sin\theta} \vspace{3mm} \cr 0
&  \cos\theta & -\displaystyle{\sin\theta  \over r} & 0 \cr }
\right), \ee with the associated metric \be
ds^2=-\left[1-{2m\left[1-\tanh\left({q^2 \over 2mR}\right)\right]
\over r}\right]dt^2-2\sqrt{2m\left[1-\tanh\left({q^2 \over
2mR}\right)\right]  \over r}dr dt+dr^2+r^2 d\Omega^2,
 \ee
 it is to be noted that $m$ $\&$ $q$ appear in the  metric (20) and (23)
 are constant of integration that will play the role of  mass
 and charge producing the field in the calculations of  energy. Also the
anstaz of the anti-symmetric field $P_{\mu \nu}$, the nonlinear
electrodynamics source used to derive this solution and the
energy-momentum tensor have the form (18) with $R=r$.

Using the coordinate transformation \be dT=dt+{Dr \over 1-D^2r^2}
dr, \ee we can eliminate the cross term of (23) to obtain \be
ds^2=-\eta_1dT^2+{dr^2 \over \eta_1}+r^2 d\Omega^2,
 \ee
 where
$\eta_1$ is defined by (20) with $R=r$.

It is our purpose to find a general solution for
 the tetrad (15) when the stress-energy momentum tensor is not vanishing
 and has the form given  by \cite{Di}
\ba
 {T_0}^0={T_1}^1, \nonu
  {T_2}^2={T_3}^3,
 \ea
  where all the other mixed spatial components equal to zero
  \cite{Di}. Then the left hand side of the second equation of
   (16) is equal zero and we can find a solution of the
  unknown function $D$ in terms of the unknown function $B$
  in the form
\be D=\displaystyle{1 \over \left(1-\displaystyle{r B' \over B}
\right)} \sqrt{ \displaystyle{k_1 B^3 \over r^3}
  \left[1-\tanh\left({q^2 \over
2mr}\right)\right] + \displaystyle{B B' \over r}
  \left(\displaystyle{r B' \over B}  -2 \right)},
\ee where $k_1$ is a constant of integration. From the first and
third equations of (16) using (26) and (27), we get the unknown
function $A$ in the form \be A=\displaystyle{k_2 \over
\left(1-\displaystyle{r B' \over B} \right)}, \ee with $k_2$ being
another  constant of integration. The general solution (27) and
(28) satisfy the differential  equations (16).

The line-element squared of (15) takes the form \be
ds^2=-{(B^2-D^2 r^2) \over A^2B^2} dt^2-{2Dr \over AB^2}dr dt+ {1
\over B^2} (dr^2+r^2 d\Omega^2). \ee  We assume $B(r)$ to be
nonvanishing so that the surface area of the sphere of a constant
${\it r}$ be finite. We also assume that $A(r)$ and $B(r)$ satisfy
the asymptotic condition,
 $\lim_{r \to \infty} A(r)$=$\lim_{r \to \infty} B(r)=1$ and
 $\lim_{r \to \infty} rB'=0$. Then, we can show from  (27), (28) and (29)
 that\\
{\bf (1)} $k_2=1$,\ \ \ {\bf (2)} \, $B(r)>0$,\ \ \  {\bf (3)} \,
$\lim_{r \to \infty} rD(r)=0,$ \quad
and \\
{\bf (4)} if $B-r B'$ vanishes at some point, then
$\left(1-\displaystyle{ B(r) k_1 \left[1-\tanh\left({q^2 \over
2mr}\right)\right] \over r } \right)<0 $ at that point.

Using the coordinate transformation \\
\be dT=dt+{ADr \over B^2-D^2r^2}dr, \ee we can eliminate the cross
term of (29) to obtain \ba ds^2= -\eta_2 dT^2 +{1 \over \eta_2}
{dr^2 \over A^2B^2} + {r^2 \over B^2}d\Omega^2 \ea with
$\eta_2={(B^2-D^2 r^2)/A^2B^2}$. Taking the new radial coordinate
$R={r/B}$, we finally get \be ds^2= -\eta_2 dT^2 +{dR^2 \over
\eta_2} +R^2d\Omega^2, \ee where \be \eta_2(R)=\left( 1-{k_1
\left[1-\tanh\left({q^2 \over 2mR}\right)\right]  \over R}
\right)\cong 1-\displaystyle{k_1 \over R}+\displaystyle{q^2 \over
R^2}. \ee Then, (32) coincides with the charged regular black hole
solution given before \cite{BI} with the mass, $m= {k_1/2}$, and
hence the general solution in the case of the spherically
symmetric tetrad when the stress-energy momentum tensor is
nonvanishing gives no more than the regular charged black hole
solution when $1-{rB'/B}$ has no zero and ${\it R}$ is
monotonically increasing function of ${\it r}$. If $1-{rB'/B}$ has
zeroes, the line-element (29) is singular at these zeroes which
lie inside the event horizon as is seen from the property (4)
mentioned above. We shall study in the future whether this
singularity at zero-points of $1-{rB'/B}$ is physically acceptable
or not.

 After using the above transformations, the tetrad (15) can be put in the form
 \be
\left({e_i}^\mu \right)= \left( \matrix{ \displaystyle{{\cal A}
\over 1-{\cal D}^2 R^2} & {\cal D} R(1-R{\cal B}') & 0 & 0
\vspace{3mm} \cr \displaystyle{{\cal A} {\cal D} R  \sin\theta
\cos\phi \over 1-{\cal D}^2R^2} &(1-R {\cal B}') \sin\theta
\cos\phi & \displaystyle{\cos\theta \cos\phi \over R}
 & -\displaystyle{\sin\phi \over R \sin\theta} \vspace{3mm} \cr
\displaystyle{{\cal A} {\cal D} R  \sin\theta \sin\phi  \over
1-{\cal D}^2 R^2} & (1-R{\cal B}') \sin\theta \sin\phi &
 \displaystyle{\cos\theta \sin\phi \over R}
 & \displaystyle{\cos\phi \over R \sin\theta} \vspace{3mm} \cr
\displaystyle{{\cal A} {\cal D} R \cos\theta  \over 1-{\cal
D}^2R^2} & (1-R{\cal B}') \cos\theta & \displaystyle{-\sin\theta
\over R} & 0 \cr } \right). \ee Here ${\cal A}$ and ${\cal D}$ are
given in terms of the unknown function $B(R)$ as \be
 {\cal A(R)} = \displaystyle{1 \over 1-R {\cal B}'}, \qquad \qquad
{\cal D(R)} =\displaystyle{1 \over 1-R {\cal B}'}
\sqrt{\displaystyle{2m\left[1-\tanh\left({q^2 \over
2mr}\right)\right]  \over R^3} + \displaystyle{{\cal B}' \over R}
\left(R {\cal B}' -2 \right)}, \ee where ${\cal
B}'=\displaystyle{d{\cal B(R)} \over dR}$. It is of interest to
note that the general solution (35) satisfies the field equations
 (10) when the anstaz
of the anti-symmetric field $P_{\mu \nu}$, the nonlinear
electrodynamics source and the energy-momentum tensor have the
form (18).

The previously obtained solutions can be verified as  special
cases of the general solution (35). The choice \be {\cal B(R)}=1,
\ee reproduces solution (21). On the other hand, the choice \be
{\cal B(R)}=\int{{1 \over R}\left(1-\sqrt{1-\displaystyle{2m(1-
\tanh\left({q^2 \over 2mR}\right)) \over R}}\right)}dR, \ee
 reproduces solution
(17). It is of interest to note that if $q=0$  then the general
solution (35) reduces to that obtained before by Mikhail et al.
\cite{MWLH} and the two choices (37) and (38) will give the
Schwarzschild solution in its standard form.

\newsection{Energy associated with the third solution}

  To make the consequence of calculations more clear,  let
us calculate the energy associated with solution (35).

 The superpotential is given by  \cite{MWHL}
 \be {{\cal U}_\mu}^{\nu \lambda} ={(-g)^{1/2} \over
2 \kappa} {P_{\chi \rho \sigma}}^{\tau \nu \lambda}
\left[\Phi^\rho g^{\sigma \chi} g_{\mu \tau}
 -\lambda g_{\tau \mu} \gamma^{\chi \rho \sigma}
-(1-2 \lambda) g_{\tau \mu} \gamma^{\sigma \rho \chi}\right], \ee
where ${P_{\chi \rho \sigma}}^{\tau \nu \lambda}$ is \be {P_{\chi
\rho \sigma}}^{\tau \nu \lambda} \stackrel{\rm def.}{=}
{{\delta}_\chi}^\tau {g_{\rho \sigma}}^{\nu \lambda}+
{{\delta}_\rho}^\tau {g_{\sigma \chi}}^{\nu \lambda}-
{{\delta}_\sigma}^\tau {g_{\chi \rho}}^{\nu \lambda} \ee with
${g_{\rho \sigma}}^{\nu \lambda}$ being a tensor defined by \be
{g_{\rho \sigma}}^{\nu \lambda} \stackrel{\rm def.}{=}
{\delta_\rho}^\nu {\delta_\sigma}^\lambda- {\delta_\sigma}^\nu
{\delta_\rho}^\lambda. \ee The energy is expressed by the surface
integral \cite{Mo2} \be E=\lim_{r \rightarrow
\infty}\int_{r=constant} {{\cal U}_0}^{0 \alpha} n_\alpha dS, \ee
where $n_\alpha$ is the unit 3-vector normal to the surface
element ${\it dS}$.

Now we are in a position to calculate the energy associated with
solution (35) using the superpotential (39). It is
 clear from (41) that, the only components which contributes to the energy is ${{\cal U}_0}^{0
 \alpha}$. Thus substituting from solution (35) into
 (39) we obtain the following non-vanishing value
 \be
{{\cal U}_0}^{0 \alpha}={2X^\alpha \over \kappa R^3}\left(2m-2m
 \tanh{\left(q^2 \over 2mR\right)}-R^2{\cal B(R)}'\right).
 \ee
 Substituting from (42) into
(41) we get \be E(R)=2m\left[1- \tanh{\left(q^2 \over
2mR\right)}\right] -R^2{\cal B(R)}', \ee which is depends on the
arbitrary function ${\cal B(R)}$.
\newsection{Regularized expression for the gravitational energy-momentum}

An important property of the tetrad fields that satisfy the
condition \be e_{i \mu}\cong \eta_{i \mu}+1/2h_{i \mu}(1/r),\ee is
that in the flat space-time limit
${e^i}_\mu(t,x,y,z)={\delta^i}_\mu$, and therefore the torsion
tensor  defined by
 \be {T^\lambda}_{\mu \nu}\stackrel{\rm
def.}{=}{e_a}^\lambda{T^a}_{\mu \nu}={\Gamma^\lambda}_{\mu
\nu}-{\Gamma^\lambda}_{\nu \mu},\ee  is vanishing, i.e.,
${T^\lambda}_{\mu \nu}=0$.  Hence for the flat space-time it is
normally to consider a set of tetrad fields such that
${T^\lambda}_{\mu \nu}=0$ {\it in any coordinate system}. However,
in general an arbitrary set of tetrad fields that yields the
metric tensor for the asymptotically flat space-time does not
satisfy the asymptotic condition given by (44). Moreover for such
tetrad fields the torsion ${T^\lambda}_{\mu \nu} \neq 0$ for the
flat space-time \cite{MRTC}$\sim$\cite{MUF}. It might be argued,
therefore, that the {\it expression for the  energy given by (41)
is restricted to particular class of tetrad fields, namely, to the
class of frames such that ${T^\lambda}_{\mu \nu}=0$ if ${e^i}_\mu$
represents the flat space-time tetrad field} \cite{MVR}. To
explain this, let us calculate the flat space-time tetrad field of
(34) using (35) which is given by \be \left({E_i}^\mu \right)
=\left(\matrix {(1-R{\cal B}') &\sqrt{R^2{\cal B}'^2-2R{\cal B}'}
&0 &0 \vspace{3mm} \cr \sqrt{R^2{\cal B}'^2-2R{\cal B}'}
\sin\theta \cos\phi    &(1-R {\cal B}') \sin\theta \cos\phi&
\displaystyle{\cos\theta \cos\phi \over R} &\displaystyle{-
\sin\phi \over R\sin\theta} \vspace{3mm} \cr \sqrt{R^2{\cal
B}'^2-2R{\cal B}'} \sin\theta \sin\phi   &(1-R {\cal B}')
\sin\theta \sin\phi& \displaystyle{\cos\theta \sin\phi \over R}
&\displaystyle{ \cos\phi \over R\sin\theta} \vspace{3mm} \cr
\sqrt{R^2{\cal B}'^2-2R{\cal B}'} \cos\theta  &(1-R {\cal B}')
\cos\theta& \displaystyle{\sin\theta \over R} & 0
 \cr } \right). \ee Expression (46) yields
the following non-vanishing torsion components: \ba \A \A
T_{001}={\cal B}', \qquad T_{112}=-r\cos(\theta)\cos\phi {\cal
B}', \qquad T_{113}=\sin(\theta)\sin\phi {\cal B}', \qquad
T_{114}=-\displaystyle{\sin(\theta)\cos\phi \sqrt{{\cal
B}'}(1-{\cal B}') \over \sqrt{R^2{\cal B}'-2R}},\nonu
\A \A T_{124}=\cos\theta \cos\phi \sqrt{R^2{\cal B}'^2-2R{\cal
B}'}, \qquad T_{134}=-\sin(\theta)\sin\phi \sqrt{R^2{\cal
B}'^2-2R{\cal B}'}, \qquad T_{212}=-R\cos(\theta)\sin\phi {{\cal
B}'},\nonu
\A \A  T_{213}=-R\sin(\theta)\cos\phi {{\cal B}'}, \qquad
T_{214}=-\displaystyle{\sin\theta \sin\phi \sqrt{{\cal
B}'}(1-{\cal B}') \over \sqrt{R^2{\cal B}'-2R}},\quad
T_{224}=\cos(\theta)\sin\phi \sqrt{R^2{\cal B}'^2-2R{\cal
B}'},\nonu
\A \A  T_{234}=\sin(\theta)\cos\phi \sqrt{R^2{\cal B}'^2-2R{\cal
B}'}, \qquad T_{312}=R\sin(\theta){{\cal B}'}, \quad
T_{314}=-\displaystyle{\cos(\theta) \sqrt{{\cal B}'}(1-{\cal B}')
\over \sqrt{R^2{\cal B}'-2R}},\nonu
\A \A  T_{324}=-\sin(\theta)\sqrt{R^2{\cal B}'^2-2R{\cal B}'}.\ea
The tetrad field (46) when written in the Cartesian coordinate
will have the form
 \be \left({E_i}^\mu(t,x,y,z) \right) =\left(\matrix
{1-R{\cal B}'& n^a \sqrt{R^2{\cal B}'^2-2R{\cal B}'} \vspace{3mm}
\cr n^\alpha  \sqrt{R^2{\cal B}'-2R{\cal B}'}&
{\delta_a}^\alpha-n_a n^\alpha  R{\cal B}' \cr } \right). \ee In
view of the geometric structure of (48), we see that, Equation
(34) using (35) do not display the asymptotic behavior required by
(44). Moreover, in general the tetrad field (48) is adapted to
accelerated observers \cite{MRTC}$\sim$\cite{MUF}. To explain
this, let us consider a boost in the x-direction of Eq. (48).  We
find \be \left({E^i}_\mu(t,x,y,z) \right) =\left(\matrix {
{vx\sqrt{R{\cal B}'^2-2{\cal B}'} +\sqrt{R}(1-R{\cal B}') \over
\gamma ^{3/2} \sqrt{R} }&{x\sqrt{R{\cal B}'^2-2{\cal B}'}
+v\sqrt{R}(1-R{\cal B}') \over \gamma ^{3/2}\sqrt{R}} & {y
\sqrt{R{\cal B}'^2-2{\cal B}'} \over R}&{z \sqrt{R{\cal
B}'^2-2{\cal B}'} \over R} \vspace{3mm} \cr {x\sqrt{R{\cal
B}'^2-2R{\cal B}'} +v\sqrt{R}(R-x^2{\cal B}') \over (\gamma
R)^{3/2}} & {xv\sqrt{R{\cal B}'^2-2R{\cal B}'}
+\sqrt{R}(R-x^2{\cal B}') \over (\gamma R)^{3/2}}& -{xy{\cal B}'
\over R} & -{xz{\cal B}' \over R} \vspace{3mm} \cr y{\sqrt{R{\cal
B}'^2-2R{\cal B}'} -vx\sqrt{R}{\cal B}' \over (R\gamma)^{3/2}} &
y{v\sqrt{R{\cal B}'^2-2R{\cal B}'} -x\sqrt{R}{\cal B}') \over
(\gamma R) ^{3/2}}&{1-y^2{\cal B}' \over R}& -{yz{\cal B} \over
R}\vspace{3mm} \cr z{\sqrt{R{\cal B}'^2-2R{\cal B}'}
-vx\sqrt{R}{\cal B}' \over ( \gamma R)^{3/2}} &z{v\sqrt{R{\cal
B}'^2-2R{\cal B}'} -x\sqrt{R}{\cal B}' \over (\gamma R)^{3/2}}
&-{zy{\cal B}' \over R} &{1-z^2{\cal B}' \over R} \cr } \right),
\ee where $v$ is the speed of the observer and
$\gamma=\sqrt{1-v^2}$. It can be shown that along an observer's
trajectory whose velocity is determined by \ba {\it
u}^\mu=E_{0}^\mu=\A \A \Biggl({vx\sqrt{R{\cal B}'^2-2{\cal B}'}
+\sqrt{R}(1-R{\cal B}') \over \gamma ^{3/2} \sqrt{R}
},{x\sqrt{R{\cal B}'^2-2{\cal B}'} +v\sqrt{R}(1-R{\cal B}') \over
\gamma ^{3/2}\sqrt{R}} ,{y \sqrt{R{\cal B}'^2-2{\cal B}'} \over
R},\nonu
\A \A {z \sqrt{R{\cal B}'^2-2{\cal B}'} \over R} \Biggr), \quad
the \quad quantities \quad {\phi_{a}}^{b}=u^\alpha
\left({E^{b}}_\beta
\partial_\alpha {E_{a}}^\beta \right),\ea constructed out from (50) are
non vanishing. This fact indicates that along the observer's path
the spatial axis ${E_{i}}^\mu$ rotate \cite{MRTC}$\sim$\cite{MUF}.
In spite of the above problems discussed for the tetrad field of
Eq. (34) using (35)  yield a satisfactory value for the total
gravitational energy-momentum, as we will discussed.

Maluf et al. \cite{MRTC}$\sim$\cite{MUF} discussed the above
problems in the framework of TEGR and constructed a regularized
expression for the gravitational energy-momentum in this frame.
They checked this expression for a tetrad field that suffer from
the above problems and obtain a very satisfactory results
\cite{MVR}. In this section we will follow the same procedure to
derive a regularized expression for the gravitational
energy-momentum defined by Eq. (41). It can be shown that the
gravitational energy-momentum contained within an arbitrary volume
$V$ of the three-dimensional spacelike hypersurface  has the form
\cite{MWHL,Mo8} \be P_\mu=\int_V d^3 x \partial_\alpha  {{\cal
U}_\mu}^{0 \alpha},\ee where
 ${{\cal U}_\mu}^{\nu \lambda}$ is given by Eq. (38).
 Expression (51) bears no relationship to the ADM
 energy-momentum \cite{MR}. $P_\mu$ transforms as a vector under the
 global SO(3,1) group. It describes the gravitational
 energy-momentum with respect to observers adapted to ${e^i}_\mu$.
 These observers are characterized by the velocity field ${\it
 u}^\mu={e_{0}}^\mu$ and by the acceleration $f^\mu$ given by
 \cite{MVR,MR}
 \be f^\mu=\displaystyle{D u^u \over
 ds}=\displaystyle{D{e_{0}}^\mu \over
 ds}=u^a\nabla_a{e_{0}}^\mu.\ee

 Our assumption is that the space-time be asymptotically flat. In
 this case the total gravitational energy-momentum is given by
 \be P_\mu=\oint_{S\rightarrow \infty} dS_\alpha \ {{\cal U}_\mu}^{0 \alpha}.\ee The
 field quantities are evaluated on a surface $S$ in the limit
 $r\rightarrow \infty$.

 In Eqs. (51) and (53) it is implicitly assumed that the reference space is determined
 by a set of tetrad fields ${e^i}_\mu$ for flat space-time such
 that the condition ${T^\lambda}_{\mu \nu}=0$ is satisfied. However, in
 general there exist flat space-time tetrad fields for which ${T^a}_{\mu \nu} \neq
 0$. In this case Eq. (51) may be generalized \cite{MVR,MR} by
 adding a suitable reference space subtraction term, exactly like
 in the Brown-York formalism \cite{BHS,YB}.

 We will denote ${T^a}_{\mu \nu}(E)=\partial_\mu {E^a}_\nu-\partial_\nu
 {E^a}_\mu$ and ${{\cal U}_\mu}^{0 \alpha} (E)$ as the expression of
 ${{\cal U}_\mu}^{0 \alpha} $
 constructed out of the flat tetrad ${E^i}_\mu$. {\it The
 regularized form of the gravitational energy-momentum $P_\mu$ is
 defined by}
 \be P_\mu=\int_{V} d^3x \partial_\alpha \left[ {{\cal U}_\mu}^{0 \alpha}(e)-
 {{\cal U}_\mu}^{0 \alpha} (E)\right].\ee
 This condition guarantees that the energy-momentum of
 the flat space-time always vanishes. The reference space-time is
 determined by tetrad fields ${E^i}_\mu$, obtained from
 ${e^i}_\mu$ by requiring the vanishing of the physical parameters
 like mass, angular momentum, etc. Assuming that the space-time is
 asymptotically flat then Eq. (54) can have the form

\be P_\mu=\oint_{S\rightarrow \infty} dS_\alpha \left[ {{\cal
U}_\mu}^{0 \alpha} (e)-{{\cal U}_\mu}^{0 \alpha} (E) \right],\ee
where the surface $S$ is established at spacelike infinity. Eq.
(55) transforms as a vector under the global SO(3,1) group
\cite{Mo8}. Now we are in a position to proof that the tetrad
field (34) sing (35) yield a satisfactory value for the total
gravitational energy-momentum.

We will integrate Eq. (55) over a surface of constant radius
$x^1=R$ and require $R\rightarrow \infty$. Therefore, the index
$\alpha$ in (55) takes the value $\alpha=1$. We need to calculate
the quantity ${{\cal U}_0}^{0 1}$ which has the form \be {{\cal
U}_0}^{0 1} (e)\cong -\displaystyle{1 \over
4\pi}R\sin(\theta)\left[{2m \over R}\left\{1-\tanh{\left(q^2 \over
2mR\right)}\right\}-R{\cal B}'+R^2{\cal B}'^2\right],\ee and the
expression of ${{\cal U}_0}^{0 1} (E)$ is obtained by just making
$m=0$ in Eq.(56). It is given by \be {{\cal U}_0}^{0 1}
(E)\cong-\displaystyle{1 \over 4\pi}R\sin(\theta)(R^2{\cal
B}'^2-R{\cal B}').\ee Thus the gravitational energy contained
within a sphere of radius $R_1$ is given by   \ba P_0 \A \A \cong
\int_{R\rightarrow R_1} d\theta d\phi \displaystyle{1 \over
4\pi}\sin(\theta)\left\{-R\left({2m \over
R}\left[1-\tanh{\left(q^2 \over 2mR\right)}\right]-R{\cal
B}'+R^2{\cal B}'^2\right)+(R^3{\cal B}'^2-R^2{\cal
B}')\right\}\nonu \A\A=2m\left[1-tanh{\left(q^2 \over 2m R_1
\right)}\right].\ea
\newsection{Main results and Discussion}

We have studied the charged solutions  in the tetrad theory of
gravitation. The axial vector part of the torsion, $a^\mu$ for
these solutions is identically vanishing and the theory reduces to
teleparallel  equivalent of general relativity coupled to
nonlinear electrodynamics.

Three different exact analytic solutions of the  field equations
are obtained for the case of spherical symmetry. These solutions
give rise to the same Riemannian metric spacetime, i.e., "regular
charged spherically symmetric spacetime". The exact solutions
(17), (21) and (35) represent a regular charged which contain the
Reissner-Nordstr$\ddot{o}$m black hole. Solution (35) is a general
solution that we can generate from it the other solutions (17) and
(21) by giving the arbitrary function ${\cal B}(r)$ the values
(36) $\&$ (37).

 It was shown by M\o ller \cite{Mo3}
that a tetrad description of a gravitational field equation allows
a more satisfactory treatment of the energy-momentum complex than
does general relativity. Therefore, we have applied the
superpotential method given by Mikhail et al.\cite{MWHL} to
calculate the energy of the central gravitating body. Calculating
the energy associated with solution (35) we obtain the formula
(44) which depends on the arbitrary function ${\cal B(R)}$.
 This formula is not accepted in physics because it makes the
 energy depends on the arbitrary function.

Maluf et al. \cite{MVR}$\sim$\cite{MUF} have derived a simple
expression for the energy-momentum flux of the gravitational
field. This expression is obtained on the assumption that Eq.(36)
represent the energy-momentum of the gravitational field on a
volume $V$ of the three-dimensional spacelike hypersurface. They
\cite{MVR,MR} gave this definition for the gravitational
energy-momentum tensor in the framework of TEGR, which require
${T^\lambda}_{\mu \nu}(E)=0$ for the flat space-time. They
extended this definition to the case where the flat space-time
tetrad fields ${E^a}_\mu$ yield ${T^\lambda}_{\mu \nu}(E) \neq 0$.
They show that \cite{MR} in the context of the regularized
gravitational energy-momentum definition it is not strictly
necessary to stipulate asymptotic boundary conditions for tetrad
fields that describe asymptotically flat space-times.

Following the same procedure given by Maluf et al.
\cite{MRTC,MVR,MR} to derive a regularized expression for the
gravitational energy-momentum in the framework of TEGR,  we derive
a similar expression of the regularized energy-momentum  in the
framework of tetrad theory of gravitation. Then we use definition
(45) of the torsion tensor and apply it to the tetrad field (34)
using (35) showing that the flat space-time associated with this
tetrad field has a non-vanishing torsion components Eq. (47) and
it is adapted to an accelerated observer (50). However, using Eq.
(55) and calculate all the necessary components we finally get Eq.
(58) which shows that the total gravitational mass does not depend
on the arbitrary function.

\vspace{2cm} \centerline{\large {\bf Acknowledgment}} The author
 would like to thank  Professor J.G. Pereira  Universidade Estadual
 Paulista.

\end{document}